# Robot-based tele-echography: Clinical evaluation of the TER system in abdominal aortic exploration

Thomas MARTINELLI[1], Jean-Luc BOSSON[2,6]., Luc BRESSOLLETTE[4], Franck PELISSIER[5], Eric BOIDARD[6], Jocelyne TROCCAZ[6], Philippe CINQUIN[6,3]

[1]Radiology Department, Grenoble University Hospital
[2]Clinical Center Research Grenoble University Hospital
[3]Information and informatics Department, Grenoble University Hospital
[4]Vascular Medicine Department, Brest University Hospital
[5]France Telecom R&D
[6]TIMC-IMAG laboratory (UMR CNRS-UJF)

Contact:
Thomas Martinelli
Tel: (33)612566372
Email: TMartinelli@chu-grenoble.fr
Address: SCRIM, CHU Grenoble, BP 217, 38043 Grenoble Cedex 09, France

Abstract

Objective: The TER system is a robot-based tele-echographic system allowing remote ultrasound (US) examination. The specialist moves a mock-up of the US probe at the master site and the robot reproduces the movements of the real probe, which sends back US images and force feedback. This tool could be used to perform US examinations in small health centers or from isolated sites. The objective of this study was to prove, in real conditions, the feasibility and the reliability of TER in detecting abdominal aortic and iliac aneurysms.
Methods: Fifty-eight patients were included in two centers in Brest and Grenoble, France. The remote examination was compared to the gold standard, the bedside examination, on measuring diameters of the aorta and iliac arteries, detecting and describing aneurysms, detecting atheromatosis, duration, and acceptability.
Results: All aneurysms (eight) were detected by both techniques as intramural thrombosis and extension to the iliac arteries. The interobserver correlation coefficient was 0.982 ($p<0.0001$) for aortic diameters. The rate of concordance between two operators in evaluating the atheromatosis was $84\% \pm 11_{CI\,95\%}$.
Conclusion: The TER system is a reliable, acceptable, and effective robot-based system to perform remote abdominal aortic US examination.

Introduction

Ultrasound (US) imaging is a key imaging modality because of its convenience and noninvasiveness, but it requires a specialist with the patient to perform the examination. This could be problematic for small healthcare centers or isolated locations. Tele-echography and robot-based echography could be a solution to provide US imaging capability to these locations.
A pure telemedicine project such as the EU-TeleInViVo project still requires a specialist to move the probe, although an expert can provide an opinion remotely. Another type of project automates an echographic examination using a robot[1-3]. Finally, a third category of robot-based systems enables the remote examination of patients by an expert who is not onsite[4-7]. The system presented herein comes within this third category; it includes force-feedback like the system used by Arbeille et al. This study aims to show that the TER system was clinically effective in real conditions of use.

Material and methods
1. Study population
From March 2004 to March 2005, 58 patients were included by two centers separated by 1000km. Any patient with an abdominal aortic aneurysm and/or atherosclerosis could be included. Neither the method used for diagnosis (CT-scan, US, clinical examination, etc.), nor the operator who carried out the examination, were used as criteria for inclusion. In order to be in more realistic conditions, no exclusion was made based on age or morphological type. Pregnant women were excluded. All patients signed the agreement after being clearly informed of the aim and the modalities of the protocol, in conformity with the bioethics law. The protocol received the agreement of our national Health Committee. Patients were informed and signed an informed consent form.

2. The Robot-Based Tele-Echography
The goal of this study was to evaluate the master-slave robotized system called TER on clinical points to demonstrate the feasibility, reproducibility, and sensitivity of a remote US examination. We focused on the detection of aortic abdominal aneurysm (AAA) and iliac aneurysms. This system has already been described in the literature [8-10]. It is a two-part system with a master site and a slave site. In this study, an experimental network was used.

   2.1. The master site
   The master site is the location of the physician during the examination. It is composed of a computer connected to a virtual US probe. This virtual probe is placed on a haptic control system to control the real probe. This haptic control integrates a PHANToM device (from SensAble Device Inc.), which has 6 degrees of freedom (dof) and renders 3D-force information. Position and orientation tracking of the virtual probe is performed within a workspace of 16 cm×13 cm×13 cm and with maximum a force of 6.4N. Information on the movement of the virtual probe is sent to the slave site to move the real probe. The physician watches the continuous ultrasound-image flow sent by the slave site. He can talk to the slave site and hear it (Figure 1). A webcam is also connected. Different data are visible on the screen of the master site (Figure 2): the ultrasound image, real-time force feedback information, and a virtual geometric model of the patient, views of the operator at the master site and of the examination room (sent by webcams) and information on the network status (transfer rates, connection status, etc.).

## 2.2. The slave site

The slave site is composed of the robot and the ultrasound device (Figure 3).

### 2.2.1. The robot

The slave robot is a parallel, uncoupled robot based on two independent structures dedicated to gross and fine positioning movements. Flexible cables are used to position and orient the US probe. The cables are connected to electrical motors. The translation movements on the surface of the body are controlled by four identical electrical motors connected by straps to a metal ring supporting the second parallel structure, as can be seen in Figure 4. The second parallel structure enables 3D orientation and translation along the Z axis of the US probe. Both subsystems can be controlled simultaneously.

The robot is controlled by a computer, which sends and receives data to and from the master site. A webcam and a microphone are also connected. Images from the ultrasound system are received by the computer and sent to the master site. Information on the screen of the slave site is similar to what is rendered at the master site.

### 2.2.2. The US system

Two different systems were used on each site: a TITAN (Sonosite Inc., Bothell, WA) with a 3- to 5-MHz probe was used on one site, an HDI 5000 (Philips, Eindhoven, Netherlands) with a 3- to 5-MHz probe was used on the other site.

## 2.3. The network

The TER system used a VTHD network during experiments. This is a very high-speed experimental communication network created and supported by France Telecom with data rates of 1 Gb/s between the two sites.

3. Description of examinations

A single-blind protocol was used. A physician included a patient and the second physician redid the US exam with no information on the patient. The second operator used the same US system with the same probe but with the remote TER system instead of the normal examination procedure. There was no operator present during the remote exam. A second person was present during the TER exam, at the slave site, to help set up the patient, install the robot, and use the US system (taking measures, freezing the image acquisition, adjusting image parameters). Although this person had to be familiar with the usage of the US and the TER system, he or she could not comment or give medical advice during the examination. Depending on the day, this person was a nurse, a resident, or a physician, present to monitor the patient during the procedure.

The exploration of the abdominal aorta and common iliac arteries was chosen because it provided the opportunity to explore a diffuse disease with multiple aspects (atheromatosis) and a focal disease (abdominal aortic aneurysm). Thus, qualitative and quantitative data could be evaluated.

For each examination (at the bedside and remotely), the following information was noted in the protocol register:

- Evaluation of the atheromatosis. Three stages can be noted: none, segmentary, and diffuse.

- Presence of an AAA and, if present, the maximum anteroposterior diameter and the presence or absence of intramural thrombosis.
- Anteroposterior diameter of each primitive iliac artery.
- Evaluation of the examination with three score values ranging from 0 to 100. The first two were for the operator and concerned the feasibility and the global quality of the exam. The third one was the evaluation of patient acceptance of the procedure.

At the end of the study, the right renal diameter was also recorded.
The classical method of examination (at the bedside) was used as the reference during this study.
Images of all measures were stored.
Data are represented by median and range and/or percentage. The reproducibility of quantitative data was assessed by calculating interobserver correlation coefficients and relative errors. The reproducibility of the qualitative data was assessed by calculating κ (kappa). We used the Stata 9 software from Stata Corp. for statistics.

# Results

1. Population study
   Fifty-eight patients (42 males and 16 females) were included from March 2004 to March 2005. Fifty-four exams (93.1%) were completed. The four failures were two dysfunctions of the haptic device, one connection problem with the VTHD, and one computer crash on the slave site.
   The median age was 63 years [27–83].
   The median BMI (weight/heigth²) was 24 [16.5–34.7] with 30% (16 patients) overweight (25<BMI<30) and 9% (five patients) obese.
   The risk factors for atheromatosis in our population are summarized in Table 1.

2. Aneurysms
   Eight patients presented an abdominal aortic aneurysm (AAA) among the 54 who had a complete examination, for a 15% prevalence of AAA. The average anteroposterior diameter was 6.5 cm [3–17]$_{CI\,95\%}$ and the median was 5.4 cm.
   All aneurysms were diagnosed by both methods, as were the six cases of intramural thrombosis and the four extensions to the iliac arteries.

3. Evaluation of the atheromatosis
   The conventional examinations found 14 patients without atheromatosis, 11 with segmentary lesions, and 28 with diffuse atheromatosis. The κ was 84% ±11$_{CI\,95\%}$. The mismatches were four diffuse forms graded as segmentary, two segmentary graded as normal, and one segmentary graded as diffuse.

4. Diameters
   a. Aorta
   The interobserver correlation coefficient was 0.982 ($p<0.0001$).
   The difference in measurements was less than 4 mm in 52 cases (96.3%), between 4 and 10 mm in one case (1.9%) and above 10 mm in one case (1.9%).

The median of the relative errors was 0 (Figure 5). The maximum and minimum relative errors were 34.5% and –26.1%. Of the measurement errors, 50% were less than 5% and 80% were less than 15%.

  b. Common iliac arteries

The interobserver correlation coefficient was 0.760 ($p<0.0001$). The median relative error was –0.044. The maximum and minimum relative errors were 27.5% and –64.1%.

5. Examination duration and global satisfaction

The medium duration increased from 12 min ± 7 with the classical US exam to 17 min ± 8 with the remote examination on T-test for repeated measures ($p<0.001$).
The evaluation of the global quality of the examination was 75.6 ± 15 (SD) ($p<0.01$) for the remote examination and 87 ± 12.5 (SD) ($p<0.01$) for the classical examination. The patients evaluated the acceptability at 84 ± 18 with 90 for the median. Two patients gave a score less than 50.

# Discussion

We collected clinical data on a new modality in radiology. We selected a pathology that allowed us to explore three types of echography data (qualitative, quantitative, semi-quantitative).

We achieved the remote vascular exploration in nearly every case and we were able to explore every objective for each patient. The failures occurred at the beginning of the study and were mainly caused by the immaturity of this prototype system.
The exploration was limited to the main vascular structures of the abdomen and pelvis but TER performed well in exploring an ideally placed structure such as the aorta and a more lateral structure such the common iliac artery.
An examination of the entire abdominal cavity seems to be more difficult. In fact, the robot control is more delicate on the lateral sides of the abdomen. Consequently, the exploration of kidneys, spleen, and liver could be more difficult because the probe must be placed on the patient's sides or on the thoracic wall. Another study is needed to evaluate this point.

Every abdominal aortic aneurysm was diagnosed. The remote measurements were reliable. The greatest discrepancy between the remote and classical examinations occurred for the patient with the largest aneurysm (17 cm in anteroposterior diameter). The interobserver variability was good in comparison to data from the literature[11, 12]. However, the reproducibility of the measurements was better for the aorta than for the iliac arteries but the absolute value for this difference was low.

In evaluating atheromatosis, the remote exam remained sensitive. We never had a difference greater than one level. We assume that some of the errors stemmed from the definition of levels. A fourth level between the 2$^{nd}$ and 3$^{rd}$ would have been useful.

The remote examination was slightly longer, but the duration of both types of procedure increased equally in a difficult examination situation. The practice of using the TER system seems to play an important role in the feeling of achievement of the examination. This is also a major point in the practitioner's overall satisfaction. In fact, this satisfaction decreases significantly with the remote exam, although nothing objectively proves that the remote examination is less effective than the bedside procedure.
The patients reported a very high satisfaction level. The two patients who gave a low score had an examination with a technical problem (among the four failed procedures).

The TER system can explore a large area but it may be harder to use over convex surfaces. The telemedicine systems using pure video transfers such as TeleInVivo are less appropriate for US exploration than TER because an expert is needed to carry out the exam; moreover, specifying probe motion remotely to the operator by oral orders may be difficult and nonintuitive. Another approach to robot-based tele-echography is illustrated in the Otelo system [4-6], which can explore only a small zone at a time and requires someone to move the system frequently to another zone, although this is done easily whatever the location. Having to reposition the system by hand may be an obstacle for easy clinical use.

Today the TER system remains the only procedure that has been clinically evaluated using appropriate methodology. The other studies[4-6] were designed to evaluate the feasibility of the exams but not their sensitivity and specificity in detecting a disease in real patients, whatever their weight or age or echogenicity. They compared the ability of the robotized examination to visualize organs well but never tested real efficacy. Our system showed good results on qualitative variables (atheromatosis, aneurysms, and intramural thrombosis) and quantitative (diameters) data.

# Conclusion

We have demonstrated that a remote examination with the TER system is possible as an everyday examination to detect, measure, and describe an abdominal aortic aneurysm or iliac aneurysm. It is a safe, sensitive, and reliable exploration. Compared to the bedside examination, it is as sensitive and the measurements are reliable. It is the first tele-echography system that has proved its reliability in real conditions. Another study has begun to determine whether these results can be extended to the exploration of the entire abdominal cavity in an emergency unit.

Legends :
Figure 1: The master site
Figure 2: Screen of the master site
Figure 3: The slave site
Figure 4: The probe and the support
Figure 5: Relative errors of the aorta diameters
Table 1: Risk factors

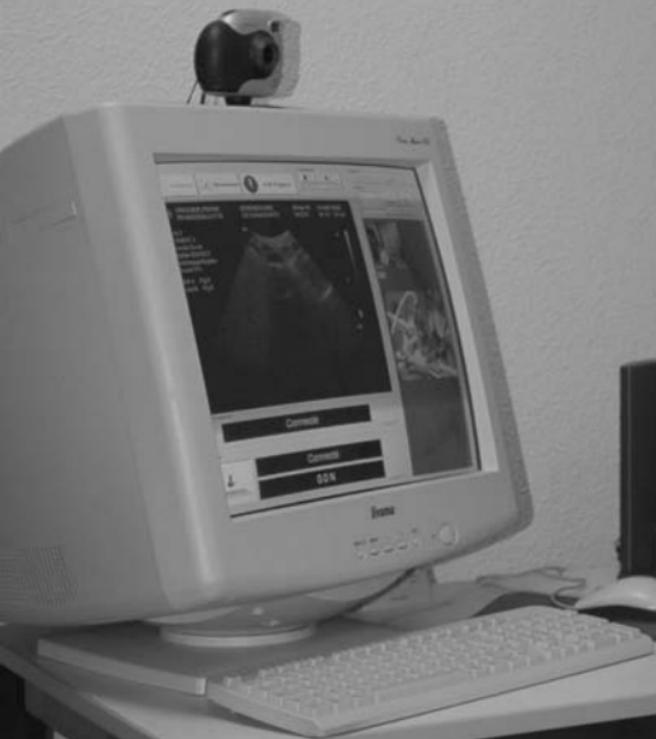

| Examen | | | Enregistrements | | | | Applications | | | |
|---|---|---|---|---|---|---|---|---|---|---|
| Connexion | Déconnexion | Arrêt d'urgence | ● | ‖ | ■ | ⏏ | Paramètres | Statistiques | Avancé... | Suite |

Conférence

Démarrer / Stopper
Ajouter / Supprimer

```
 DUP , PI                               2004 Mai 19 11:56
Gén                                                  Abd
                                                     C60
                                                     ▪▪▪▪
                                                     CF
                                                   139
                                                    IM
                                                   0,4

                                                    14
```

| 2D | Gén | G/D | H/B | 8 | Biops. | Double |

**Image échographique**

Connecté

**Haptique**

Connecté

0.0 N

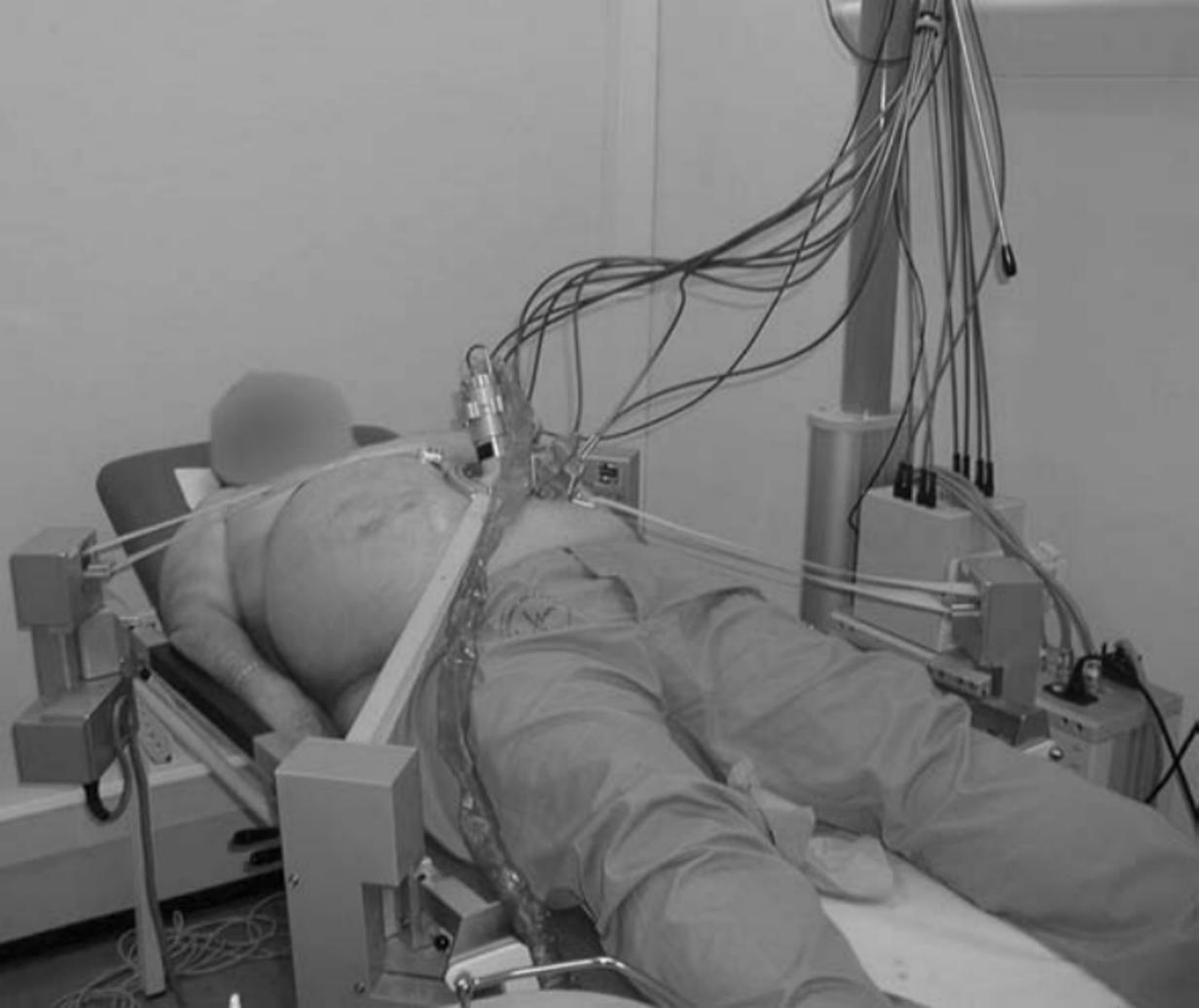

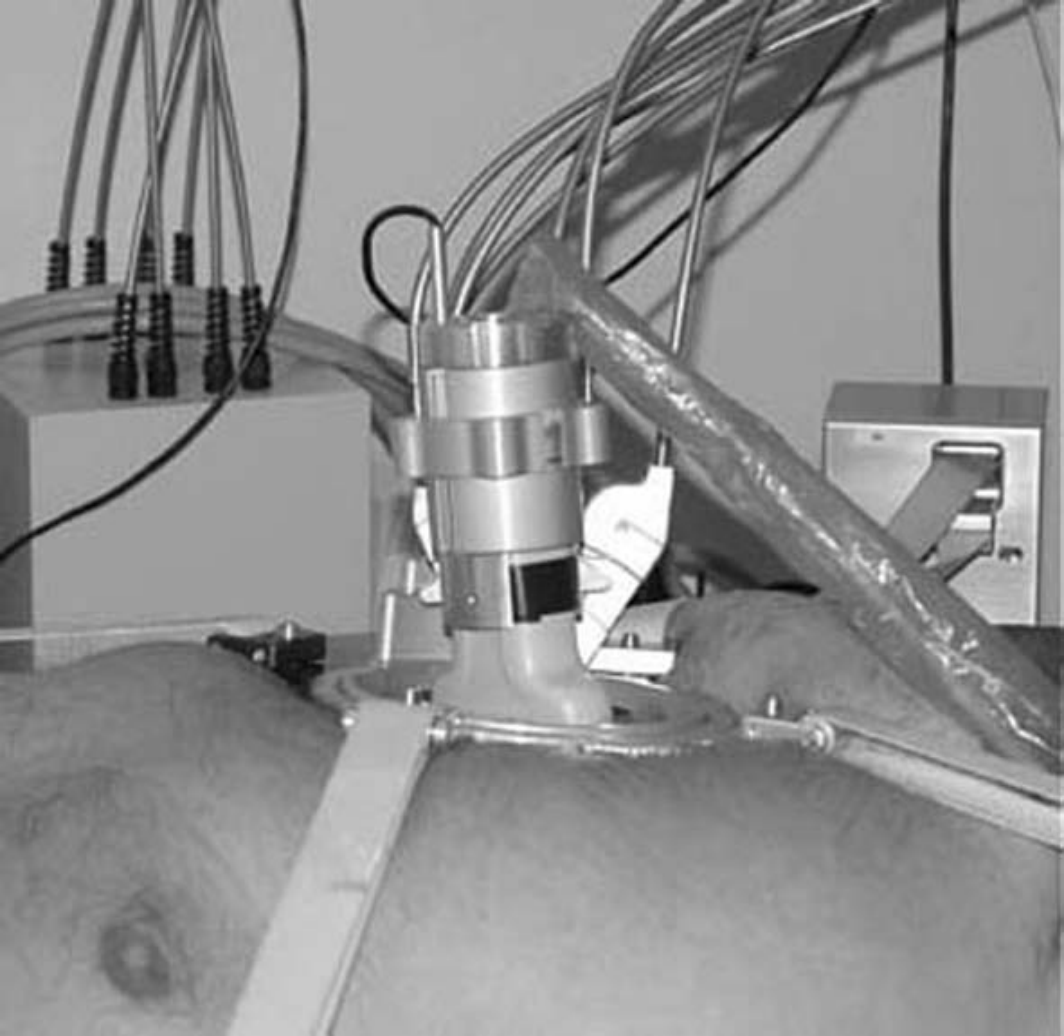

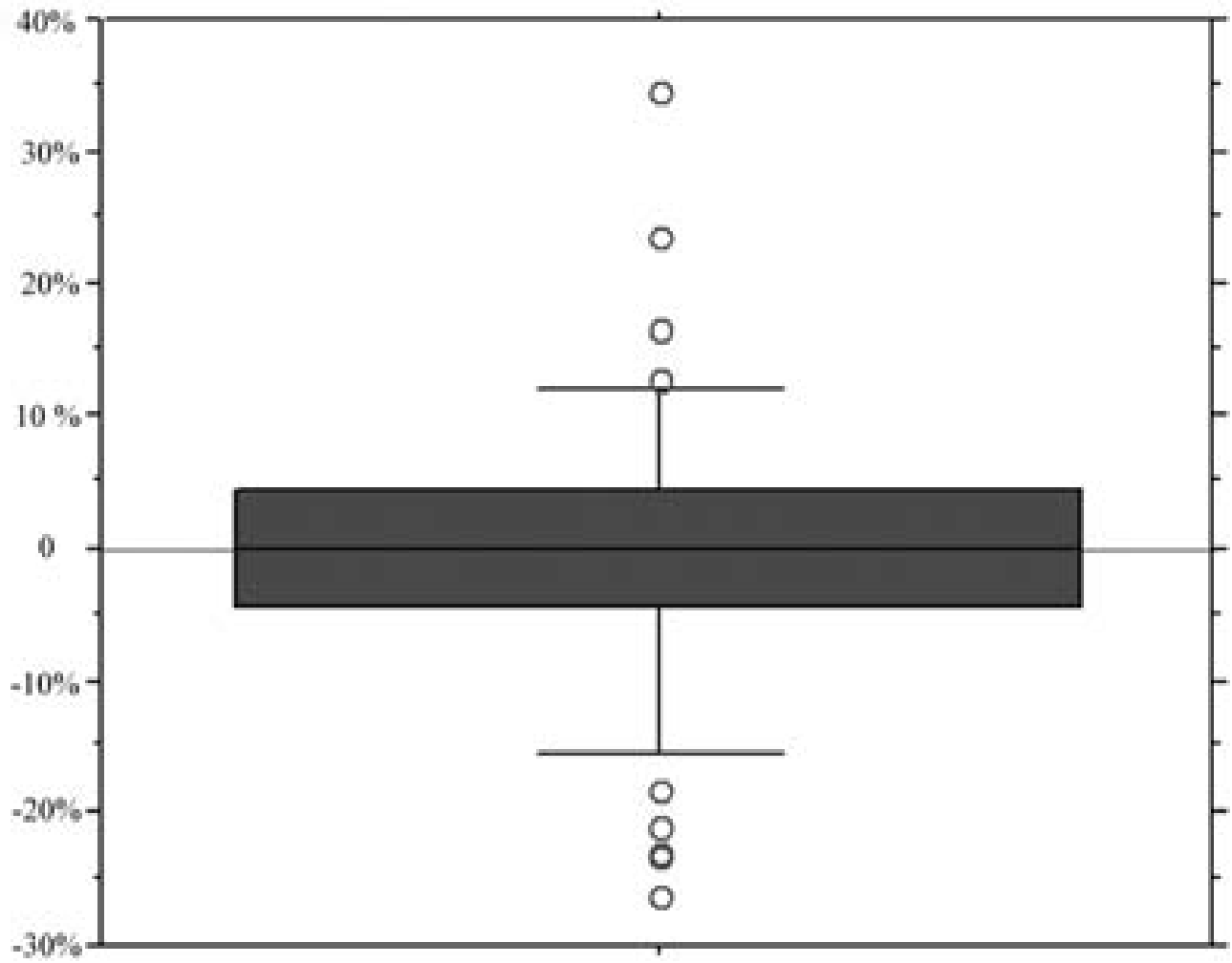

| Risk Factor | Nb of patients |
|---|---|
| Hypertension | 14 (26%) |
| Dyslipemia | 9 (17%) |
| Smoker (old or active) | 30 (52%) |
| Coronary failure | 9 (17%) |
| Supra-aortic trunk atheromatosis | 8 (15%) |
| Claudication (stage II) | 13 (21%) |